\documentclass[10pt,a4paper,onecolumn]{article}
\usepackage{marginnote}
\usepackage{graphicx}
\usepackage{xcolor}
\usepackage{authblk,etoolbox}
\usepackage{titlesec}
\usepackage{calc}
\usepackage{tikz}
\usepackage{hyperref}
\hypersetup{colorlinks,breaklinks=true,
            urlcolor=[rgb]{0.0, 0.5, 1.0},
            linkcolor=[rgb]{0.0, 0.5, 1.0}}
\usepackage{caption}
\usepackage{tcolorbox}
\usepackage{amssymb,amsmath}
\usepackage{ifxetex,ifluatex}
\usepackage{seqsplit}
\usepackage{xstring}

\usepackage{float}
\let\origfigure\figure
\let\endorigfigure\endfigure
\renewenvironment{figure}[1][2] {
    \expandafter\origfigure\expandafter[H]
} {
    \endorigfigure
}

\usepackage{fixltx2e} 
\usepackage[
  backend=biber,
]{biblatex}
\bibliography{references.bib}

\let\textttOrig=\texttt
\def\texttt#1{\expandafter\textttOrig{\seqsplit{#1}}}
\renewcommand{\seqinsert}{\ifmmode
  \allowbreak
  \else\penalty6000\hspace{0pt plus 0.02em}\fi}

\makeatletter
\let\href@Orig=\href
\def\href@Urllike#1#2{\href@Orig{#1}{\begingroup
    \def\Url@String{#2}\Url@FormatString
    \endgroup}}
\def\href@Notdoi#1#2{\def\tempa{#1}\def\tempb{#2}%
  \ifx\tempa\tempb\relax\href@Urllike{#1}{#2}\else
  \href@Orig{#1}{#2}\fi}
\def\href#1#2{%
  \IfBeginWith{#1}{https://doi.org}%
  {\href@Urllike{#1}{#2}}{\href@Notdoi{#1}{#2}}}
\makeatother

\newlength{\cslhangindent}
\newlength{\csllabelwidth}
\newenvironment{CSLReferences}[3] 
 {
  \setlength{\parindent}{0pt}
  \ifodd #1 \everypar{\setlength{\hangindent}{\cslhangindent}}\ignorespaces\fi
  \ifnum #2 > 0
  \setlength{\parskip}{#2\baselineskip}
  \fi
 }%
 {}
\usepackage{calc}

\usepackage[top=3.5cm, bottom=3cm, right=1.5cm, left=1.0cm,
            headheight=2.2cm, reversemp, includemp, marginparwidth=4.5cm]{geometry}

\titleformat{\section}
  {\normalfont\sffamily\Large\bfseries}
  {}{0pt}{}
\titleformat{\subsection}
  {\normalfont\sffamily\large\bfseries}
  {}{0pt}{}
\titleformat{\subsubsection}
  {\normalfont\sffamily\bfseries}
  {}{0pt}{}
\titleformat*{\paragraph}
  {\sffamily\normalsize}

\usepackage{fancyhdr}
\makeatletter
\let\ps@plain\ps@fancy
\fancyheadoffset[L]{4.5cm}
\fancyfootoffset[L]{4.5cm}

\definecolor{linky}{rgb}{0.0, 0.5, 1.0}

\newtcolorbox{repobox}
   {colback=red, colframe=red!75!black,
     boxrule=0.5pt, arc=2pt, left=6pt, right=6pt, top=3pt, bottom=3pt}

\newcommand{\ExternalLink}{%
   \tikz[x=1.2ex, y=1.2ex, baseline=-0.05ex]{%
       \begin{scope}[x=1ex, y=1ex]
           \clip (-0.1,-0.1)
               --++ (-0, 1.2)
               --++ (0.6, 0)
               --++ (0, -0.6)
               --++ (0.6, 0)
               --++ (0, -1);
           \path[draw,
               line width = 0.5,
               rounded corners=0.5]
               (0,0) rectangle (1,1);
       \end{scope}
       \path[draw, line width = 0.5] (0.5, 0.5)
           -- (1, 1);
       \path[draw, line width = 0.5] (0.6, 1)
           -- (1, 1) -- (1, 0.6);
       }
   }

\patchcmd{\@maketitle}{center}{flushleft}{}{}
\patchcmd{\@maketitle}{center}{flushleft}{}{}
\patchcmd{\@maketitle}{\LARGE}{\LARGE\sffamily}{}{}
\def\maketitle{{%
  
  \AB@maketitle}}
\makeatletter
\renewcommand\AB@affilsepx{ \protect\Affilfont}
\renewcommand\AB@affilnote[1]{{\bfseries #1}\hspace{3pt}}
\renewcommand{\affil}[2][]%
   {\newaffiltrue\let\AB@blk@and\AB@pand
      \if\relax#1\relax\def\AB@note{\AB@thenote}\else\def\AB@note{#1}%
        \setcounter{Maxaffil}{0}\fi
        \begingroup
        \let\href=\href@Orig
        \let\texttt=\textttOrig
        \let\protect\@unexpandable@protect
        \def\thanks{\protect\thanks}\def\footnote{\protect\footnote}%
        \@temptokena=\expandafter{\AB@authors}%
        {\def\\{\protect\\\protect\Affilfont}\xdef\AB@temp{#2}}%
         \xdef\AB@authors{\the\@temptokena\AB@las\AB@au@str
         \protect\\[\affilsep]\protect\Affilfont\AB@temp}%
         \gdef\AB@las{}\gdef\AB@au@str{}%
        {\def\\{, \ignorespaces}\xdef\AB@temp{#2}}%
        \@temptokena=\expandafter{\AB@affillist}%
        \xdef\AB@affillist{\the\@temptokena \AB@affilsep
          \AB@affilnote{\AB@note}\protect\Affilfont\AB@temp}%
      \endgroup
       \let\AB@affilsep\AB@affilsepx
}
\makeatother

\renewcommand\Affilfont{\sffamily\small\mdseries}
\ifnum 0\ifxetex 1\fi\ifluatex 1\fi=0 
  \usepackage[T1]{fontenc}
  \usepackage[utf8]{inputenc}

\else 
  \ifxetex
    \usepackage{mathspec}
    \usepackage{fontspec}

  \else
    \usepackage{fontspec}
  \fi
  \defaultfontfeatures{Ligatures=TeX,Scale=MatchLowercase}

\fi
\IfFileExists{upquote.sty}{\usepackage{upquote}}{}
\IfFileExists{microtype.sty}{%
\usepackage{microtype}
\UseMicrotypeSet[protrusion]{basicmath} 
}{}

\usepackage{hyperref}
\hypersetup{unicode=true,
            pdftitle={nrCascadeSim - A simulation tool for nuclear recoil cascades resulting from neutron capture},
            pdfborder={0 0 0},
            breaklinks=true}
\let\addcontentslineOrig=\addcontentsline
\def\addcontentsline#1#2#3{\bgroup
  \let\texttt=\textttOrig\addcontentslineOrig{#1}{#2}{#3}\egroup}
\let\markbothOrig\markboth
\def\markboth#1#2{\bgroup
  \let\texttt=\textttOrig\markbothOrig{#1}{#2}\egroup}
\let\markrightOrig\markright
\def\markright#1{\bgroup
  \let\texttt=\textttOrig\markrightOrig{#1}\egroup}

\usepackage{graphicx,grffile}
\makeatletter
\def\maxwidth{\ifdim\Gin@nat@width>\linewidth\linewidth\else\Gin@nat@width\fi}
\def\maxheight{\ifdim\Gin@nat@height>\textheight\textheight\else\Gin@nat@height\fi}
\makeatother
\setkeys{Gin}{width=\maxwidth,height=\maxheight,keepaspectratio}
\IfFileExists{parskip.sty}{%
\usepackage{parskip}
}{
\setlength{\parindent}{0pt}
\setlength{\parskip}{6pt plus 2pt minus 1pt}
}
\ifx\paragraph\undefined\else
\let\oldparagraph\paragraph
\renewcommand{\paragraph}[1]{\oldparagraph{#1}\mbox{}}
\fi
\ifx\subparagraph\undefined\else
\let\oldsubparagraph\subparagraph
\renewcommand{\subparagraph}[1]{\oldsubparagraph{#1}\mbox{}}
\fi

\title{\texttt{nrCascadeSim} - A simulation tool for nuclear recoil
cascades resulting from neutron capture}

        \author[1]{A.N. Villano}
          \author[1]{Kitty Harris}
          \author[2]{Staci Brown}
    
      \affil[1]{Department of Physics, University of Colorado Denver,
Denver CO 80217, USA}
      \affil[2]{Department of Applied Mathematics \& Statistics,
University of New Mexico, Albuquerque NM 87131, USA}
  \date{\vspace{-7ex}}

\begin{document}
\maketitle

\marginpar{

  \begin{flushleft}
  \sffamily\small

  {\bfseries DOI:} \href{https://doi.org/DOI unavailable}{\color{linky}{DOI unavailable}}

  \vspace{2mm}

  {\bfseries Software}
  \begin{itemize}
    \setlength\itemsep{0em}
    \item \href{N/A}{\color{linky}{Review}} \ExternalLink
    \item \href{NO_REPOSITORY}{\color{linky}{Repository}} \ExternalLink
    \item \href{DOI unavailable}{\color{linky}{Archive}} \ExternalLink
  \end{itemize}

  \vspace{2mm}

  \par\noindent\hrulefill\par

  \vspace{2mm}

  {\bfseries Editor:} \href{https://example.com}{Pending
Editor} \ExternalLink \\
  \vspace{1mm}
    {\bfseries Reviewers:}
  \begin{itemize}
  \setlength\itemsep{0em}
    \item \href{https://github.com/Pending Reviewers}{@Pending
Reviewers}
    \end{itemize}
    \vspace{2mm}

  {\bfseries Submitted:} N/A\\
  {\bfseries Published:} N/A

  \vspace{2mm}
  {\bfseries License}\\
  Authors of papers retain copyright and release the work under a Creative Commons Attribution 4.0 International License (\href{http://creativecommons.org/licenses/by/4.0/}{\color{linky}{CC BY 4.0}}).

  \end{flushleft}
}

\hypertarget{summary}{%
\section{Summary}\label{summary}}

Neutron capture-induced nuclear recoils have emerged as an important
tool for detector calibrations in direct dark matter detection and
coherent elastic neutrino-nucleus scattering (CE\(\mathrm{\nu}\)NS).

\texttt{nrCascadeSim} is a C++ command-line tool for generating
simulation data for energy deposits resulting from neutron capture on
pure materials. Presently, capture events within silicon, germanium,
neon, and argon are supported. While the software was developed for
solid state detector calibration, it can be used for any application
which requires simulated neutron capture-induced nuclear recoil data.

A ``cascade'' occurs when a neutron becomes part of a nucleus. The
neutron can be captured to one of many discrete energy levels, or
states; if the energy level is nonzero (not the ground state), then the
state will eventually change so that it is zero. This can happen either
all at once or in multiple steps --- that is, the captured neutron may
go from its state to the ground state, or it may go to another state
with lower energy that is not the ground state (provided that one
exists). The cascade refers to the particular ``path'' of energy levels
that a captured neutron takes to get to the ground state from the
neutron separation energy. Currently, the code assumes that the neutrons
that enter the nuclear system have zero kinetic energy; this is a good
approximation for thermal neutrons because 0.0254~eV (the average
kinetic energy of a thermal neutron) is small compared to most nuclear
recoil energy scales, making it negligible.

\texttt{nrCascadeSim} models many of these cascades at once and saves
the energies along with other useful data to a single file. The output
file is a \texttt{ROOT} file (Brun \& Rademakers, 1997).

\hypertarget{models-used}{%
\section{Models Used}\label{models-used}}

When modeling deposits from neutron capture events, we want to look at
the recoil of the atom as a result of these cascades. To determine how
much energy is deposited, we must track how much the atom slows down
between steps of the cascade as well as how each nuclear state change
affects the atom's kinetic energy. \texttt{nrCascadeSim} assumes a
constant deceleration that results from the atom colliding with other
nearby electrons and nuclei. This means that it must simulate, along
with the steps of the cascade, the time between each state --- to
calculate how much the atom slows down. And it must also simulate the
angle between the atom's momentum before a decay and the momentum boost
(gamma ray) resulting from the decay --- to calculate the resulting
momenta. The time between steps is simulated as an exponential random
variable based on the state's half-life, and the angle is simulated as
having an isotropic distribution. Cascade selection is weighted by
isotope abundance (Rosman \& Taylor, 1998; Sonzogni \& Shu, 2020) and
cross-section as well as the probability of the energy level. In
existing levelfiles, energy levels are derived from (Islam et al., 1991)
for germanium and from (Raman et al., 1992) for silicon.

The above process models the recoil energies, and the output gives both
the total recoil energy for a cascade as well as the energy per step.
For some applications, this may be the desired output, or the user may
already have a particular process they will use for converting this
energy to what they wish to measure. However, we also include, for
convenience, the ionization yield and ionization energy of these
recoils. Ionization yield is a fraction that, when multiplied by the
energy, gives the ionization energy, and ionization energy is the amount
of energy that would be read out if an otherwise equivalent electron
recoil were to occur. This calculation is useful because many
solid-state detectors read out the ionization energy for nuclear
recoils. This ionization yield assumes the Lindhard model (Lindhard et
al., 1963).

\begin{figure}
\centering
\includegraphics{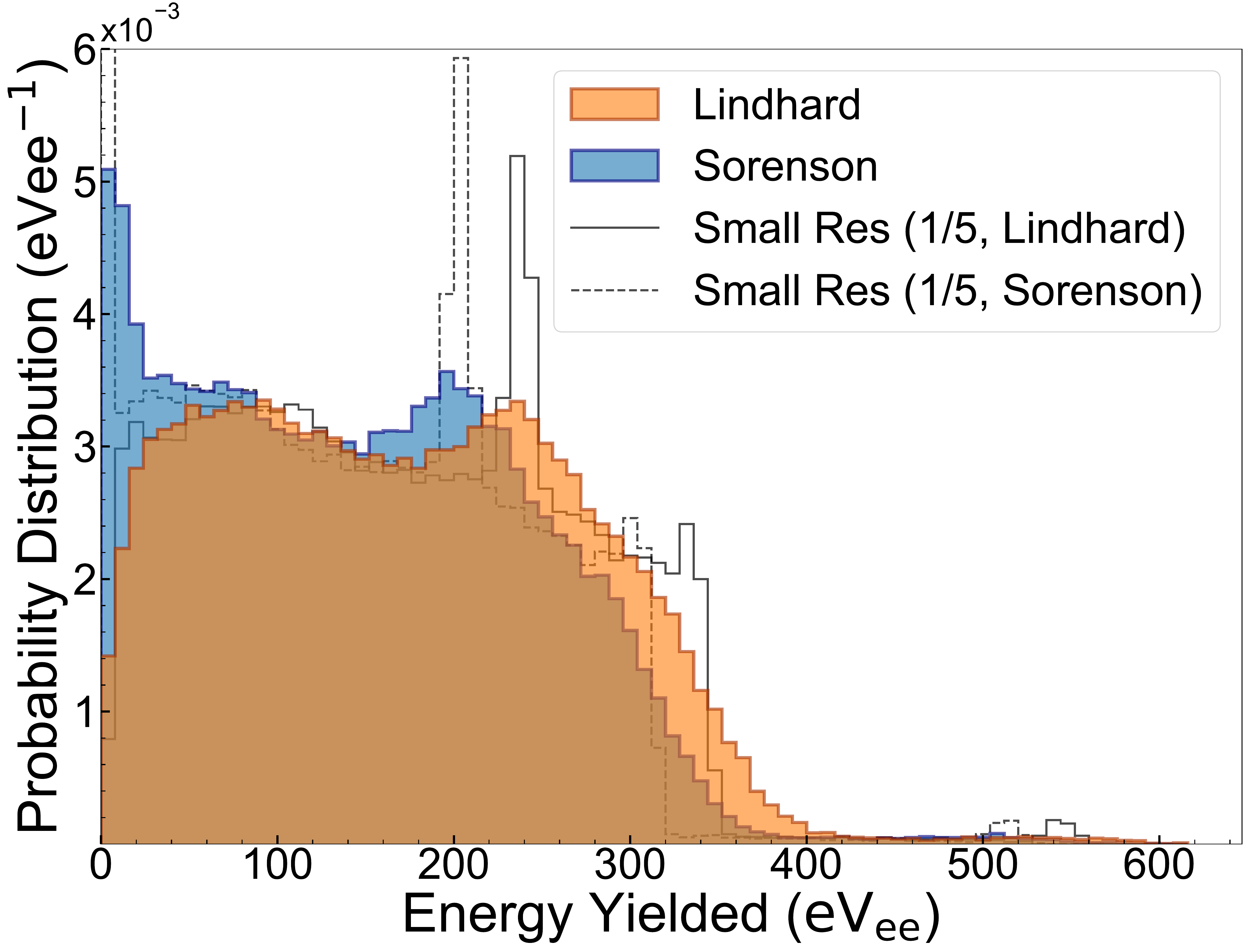}
\caption{An overlaid histogram showing an example use case in which
points are generated and then multiple yield models and resolutions are
applied. The ``Small Res (1/5)'' histograms have Gaussians with 1/5 of
the width of their counterparts. \label{LindvSor_fig}}
\end{figure}

Figure \ref{LindvSor_fig} compares the normalized frequencies of
ionization energies from the Lindhard (Lindhard et al., 1963) model with
the Sorensen (Sorensen, 2015) yield model, which is applied after the
simulation using Python, and applies detector resolution models to both.
This figure demonstrates one example of user-applied analysis utilizing
the energy deposits at each step instead of the ionization energy.

\hypertarget{statement-of-need}{%
\section{Statement of Need}\label{statement-of-need}}

The goal of this software is to simplify the computation of the nuclear
recoil spectrum following neutron capture for a variety of applications.
These include nuclear recoil calibrations for dark matter direct
detection and coherent neutrino detection (CE\(\mathrm{\nu}\)NS). In
these cases as the particle detection has become more sensitive
(detectors having a lower energy threshold) it is now possible to use
the capture-induced nuclear recoil events for detector calibrations.
Additionally, thermalized neutrons will provide large backgrounds that
have heretofore not been modeled. The key roadblock to studying these
scenarios is the complexity of calculating the nuclear recoil spectrum.

\texttt{nrCascadeSim} addresses this need by allowing users to generate
nuclear recoil simulations that reflect a variety of single-element
detector setups. The energy levels that the recoiling nuclei may pass
between and their respective lifetimes are customizable, and multiple
isotopes of the same element can be present within the same simulation.
Pre-defined energy level files exist for silicon and germanium, which
take into account the natural abundance data of each isotope in (Rosman
\& Taylor, 1998) and (Sonzogni \& Shu, 2020). Output values include
energy deposits at each step along each individual cascade, total
kinetic energy deposits, and ionization energy deposits.

\hypertarget{state-of-the-field}{%
\section{State of the Field}\label{state-of-the-field}}

\begin{figure}
\centering
\includegraphics{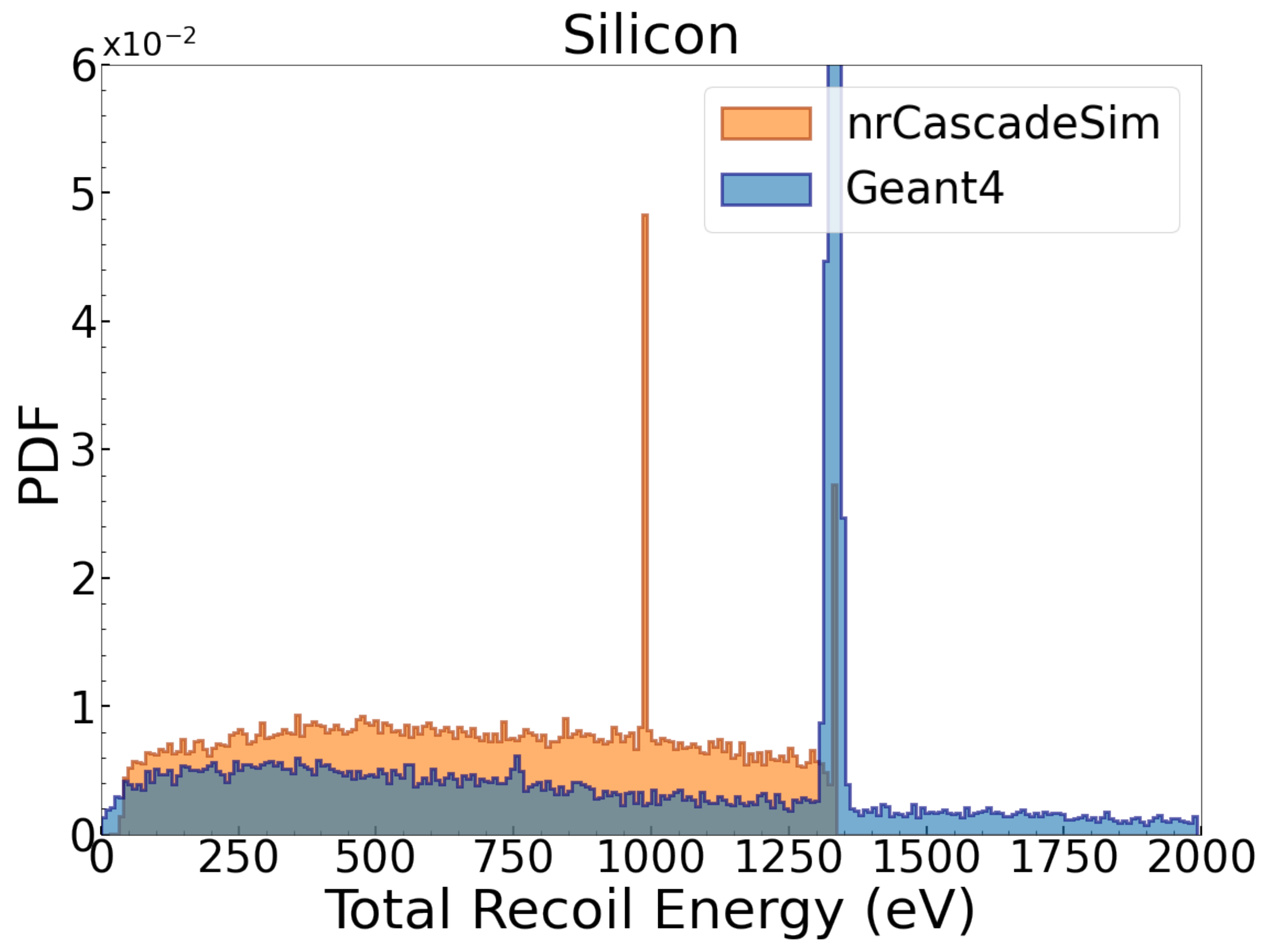}
\caption{An overlaid histogram showing how the Geant4 \texttt{v10.7.3}
energy deposits compare with those from \texttt{nrCascadeSim} for
natural silicon. \label{G4comp}}
\end{figure}

While there are tools, such as the open-source GEANT4 (Agostinelli et
al., 2003) framework, that allow users to simulate neutron capture,
existing tools are not built specifically for neutron capture-based
nuclear recoils as \texttt{nrCascadeSim} is and therefore use some
underlying assumptions that \texttt{nrCascadeSim} does not. The main
approximation often used in GEANT4 that we avoid in
\texttt{nrCascadeSim} is that all recoils decay directly to the ground
state. While this works for some applications, it is necessary to be
more precise when an accurate spectrum of neutron capture-based recoils
is needed for analyses such as calibration or background subtraction.
Figure \ref{G4comp} shows a comparison for the energy deposits produced
by Geant4 for natural silicon compared with those produced by
\texttt{nrCascadeSim}. The figure does not include any instrumentation
resolution and shows a highly prominent peak around 1.25 keV recoil
energy (coming from capture on \(^{29}\)Si directly to the ground state)
whereas the \texttt{nrCascadeSim} shows another direct-to-ground
contribution (from capture on \(^{28}\)Si) at around 1.0 keV recoil
energy and generally far more ``spread out'' recoils coming from two- or
more step cascades.

Recently, the power of the neutron capture-induced events has been
acknowledged in the CE\(\mathrm{\nu}\)NS field (Thulliez \& others,
2021). That initial study, however, used the FIFRELIN code (Litaize \&
Serot, 2010), which was originally developed for modeling fission
fragments and has been updated to use statistical models of gamma
emission for the purpose of modeling fission-fragment deexcitation
(Litaize et al., 2014). \texttt{nrCascadeSim} takes the complementary
approach of beginning with small to medium-sized nuclei and modeling the
cascades in more exact detail. The goal is for the code to be extended
to heavier nuclei but still using this detailed approach.

\hypertarget{acknowledgments}{%
\section{Acknowledgments}\label{acknowledgments}}

This material is based upon work supported by the U.S. Department of
Energy, Office of Science, Office of High Energy Physics (HEP) under
Award Number DE-SC0021364.

\hypertarget{references}{%
\section*{References}\label{references}}
\addcontentsline{toc}{section}{References}

\hypertarget{refs}{}
\begin{CSLReferences}{1}{0}
\leavevmode\hypertarget{ref-Geant4}{}%
Agostinelli, S., Allison, J., Amako, K., Apostolakis, J., Araujo, H.,
Arce, P., Asai, M., Axen, D., Banerjee, S., Barrand, G., Behner, F.,
Bellagamba, L., Boudreau, J., Broglia, L., Brunengo, A., Burkhardt, H.,
Chauvie, S., Chuma, J., Chytracek, R., \& Cooperman, G. (2003). GEANT4 -
a simulation toolkit. \emph{Nuclear Instruments and Methods in Physics
Research A}, \emph{506}, 250--303.
\url{https://doi.org/10.1016/S0168-9002(03)01368-8}

\leavevmode\hypertarget{ref-ROOT}{}%
Brun, R., \& Rademakers, F. (1997). ROOT - an object oriented data
analysis framework. \emph{Nucl. Inst. \& Meth. In Phys. Res. A},
\emph{389}, 81--86. See also "ROOT" {[}software{]}, Release v6.22/00,
02/07/2020, (No DOI available for this version, closest version linked).
\url{https://doi.org/10.5281/zenodo.3895852}

\leavevmode\hypertarget{ref-Ge}{}%
Islam, M. A., Kennett, T. J., \& Prestwich, W. V. (1991). Radiative
strength functions of germanium from thermal neutron capture.
\emph{Physical Review C}, \emph{43}, 1086--1098.
\url{https://doi.org/10.1103/physrevc.43.1086}

\leavevmode\hypertarget{ref-lindhard}{}%
Lindhard, J., Nielsen, V., \& Scharff, M. (1963). Integral equations
governing radiation effects. \emph{Mat. Fys. Medd. Dan. Vid. Selsk.},
\emph{33}, 1--41.

\leavevmode\hypertarget{ref-PhysRevC.82.054616}{}%
Litaize, O., \& Serot, O. (2010). Investigation of phenomenological
models for the monte carlo simulation of the prompt fission neutron and
\(\ensuremath{\gamma}\) emission. \emph{Phys. Rev. C}, \emph{82},
054616. \url{https://doi.org/10.1103/PhysRevC.82.054616}

\leavevmode\hypertarget{ref-FIFRELIN}{}%
Litaize, O., Serot, O., \& Berge, L. (2014).
\url{https://www.oecd-nea.org/science/meetings/pnd22/presentations/1-LITAIZE.pdf}

\leavevmode\hypertarget{ref-Si}{}%
Raman, S., Jurney, E. T., \& Lynn, J. E. (1992). Thermal-neutron capture
by silicon isotopes. \emph{Physical Review C}, \emph{46}, 972--983.
\url{https://doi.org/10.1103/PhysRevC.46.972}

\leavevmode\hypertarget{ref-abundances}{}%
Rosman, K. J. R., \& Taylor, P. D. P. (1998). Isotopic compositions of
the elements 1997. In \emph{Pure \& Appl. Chem.} (Vol. 70, pp.
217--235). International Union of Pure; Applied Chemistry.
\url{https://doi.org/10.1515/iupac.70.0026}

\leavevmode\hypertarget{ref-nudat2}{}%
Sonzogni, A., \& Shu, B. (2020). \emph{Nudat 2}. Brookhaven National
Laboratory. \url{https://www.nndc.bnl.gov/nudat2/}

\leavevmode\hypertarget{ref-sorensen}{}%
Sorensen, P. (2015). Atomic limits in the search for galactic dark
matter. \emph{Phys. Rev. D}, \emph{91}.
\url{https://doi.org/10.1103/PhysRevD.91.083509}

\leavevmode\hypertarget{ref-crab}{}%
Thulliez, L., \& others. (2021). \emph{Calibration of nuclear recoils at
the 100 {eV} scale using neutron capture}. \emph{16}(07), P07032.
\url{https://doi.org/10.1088/1748-0221/16/07/p07032}

\end{CSLReferences}

\end{document}